\documentclass[aps,prb,preprint,showpacs,showkeys,superscriptaddress,amysymb]{revtex4-1}

\usepackage{makeidx}
\usepackage{amsmath}
\usepackage{graphicx}
\usepackage{dcolumn}
\usepackage{bm}
\usepackage[usenames]{color}
\usepackage{float}
\begin{document}

\title{Exploring the transport properties of polytypic and twin-plane nanowires: from tunneling phase-time to spin-orbit interaction effects}

\author{M. Rebello Sousa Dias}
\affiliation{Departamento de F\'{\i}sica, Universidade Federal de S\~{a}o Carlos, 13565-905 S\~{a}o Carlos, SP, Brazil}

\author{L. Villegas-Lelovsky}
\address{Instituto de F\'{\i}sica, Universidade de Brasilia,70910-900, Brasilia, DF, Brazil}

\author{L. Diago-Cisneros}
\affiliation{Facultad de F\'{\i}sica, Universidad de La Habana, Cuba}

\author{L. K. Castelano}
\affiliation{Departamento de F\'{\i}sica, Universidade Federal de S\~{a}o Carlos, 13565-905 S\~{a}o Carlos, SP, Brazil}

\author{D. F. Cesar}
\affiliation{Departamento de F\'{\i}sica, Universidade Federal de S\~{a}o Carlos, 13565-905 S\~{a}o Carlos, SP, Brazil}

\author{G. E. Marques}
\affiliation{Departamento de F\'{\i}sica, Universidade Federal de S\~{a}o Carlos, 13565-905 S\~{a}o Carlos, SP, Brazil}

\author{V. Lopez-Richard}
\affiliation{Departamento de F\'{\i}sica, Universidade Federal de S\~{a}o Carlos, 13565-905 S\~{a}o Carlos, SP, Brazil}

\begin{abstract}

The variety of nanowire crystal structures gave rise to unique and novel transport phenomena. In particular, we have explored the
superlattice profile generated by strain field modulation in twin-plane nanowires for the tuning of transport channels and the
built-in spin-orbit potential profile of polytypic nanowires, in order to realize a spin filter. The Multicomponent Scattering
Approach has been used in terms of the Transfer Matrix Method to describe the phase-time of charge carriers. This system showed
advantages for attaining conditions for the propagation of wave packets with negative group velocity. Moreover, the spin transport
effect of a potential profile with volumetric spin-orbit bulk inversion asymmetry, as present on polytypic nanowires, was
described through the Reverse Runge-Kutta Method. Using the peculiar symmetry of the excited states we have characterized a dominant spin
dependence on structural parameters that results in effective spin filtering.
\end{abstract}

\date{\today}
\maketitle

\section{introduction}

\begin{figure}
\includegraphics[scale=0.8]{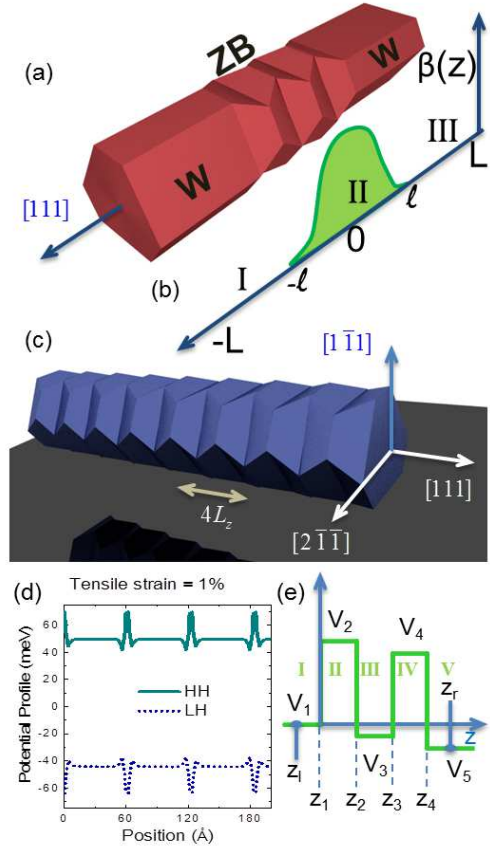}
\caption{(a) Schematic diagram of a polytypic nanowire
(PNW). (b) Electronic structure profile of a PNW used in the simulation of the spin transport properties, where $\beta(z)$ is the
Dresselhaus or BIA spin-orbit potentials. (c) Schematic diagram of a twin-plane nanowire (TPNW) used in our simulation for the electronic structure. (d) Electronic structure profile of the top-most valence subbands. (e) Notation used to characterize the scattering potential profile. The segment
$[z_{L},z_{R}]$ represents the periodic cell~\cite{SADiago2012} for the traveling modes. } \label{Fig1}
\end{figure}

The integration of nanowire (NW) architecture as a building-block for electronic and photonic applications has become an issue of
considerable research interest.~\cite{Samuelson1, Lieber} Therefore, the study of their transport properties is a central task for both
theoretical and experimental endeavors boosted by the developments of growth techniques that allow minute electronic structure engineering.
The ability of thorough control of NW crystal structures by varying parameters as growth temperature and pressure, NW diameter and surface
density, precursor molar fraction, III/V aspect ratio, and incorporation of impurity atoms allows the fabrication of a variety of high
quality wires displaying zinc-blende, twin-planes (Fig.~\ref{Fig1}(a)), stacking faults (Fig.~\ref{Fig1}(d)), and  wurtzite crystalline
structures.~\cite{Bakkers, Samuelson, Samuelson2, InSbpol} The improving of this level of control can lead to a variety of electronic behaviors
and these new prospects have opened different opportunities for the tuning of their transport properties.

The effect of volumetric spin-orbit (SO) bulk inversion asymmetry
(BIA) in wires with zincblende crystalline structure has been previously
discussed,~\cite{Lelovsky} where the degeneracy of spin states for each Bloch
wavevector $k$, in the absence of an inversion center, is broken. This
leads to the spin splitting of energy states of both electrons and holes
without lifting the Kramer degeneracy in the absence of an external
magnetic field. It was shown that, although the
ground state of a cylindrical NW is always doubly
degenerate in terms of the spin polarization at any value of $k_z$,
regardless of the direction,
different terms induced by the Dresselhaus SO contribution, split excited
levels when considering
terms proportional to $k_z$, the longitudinal component of the wavevector. This is a dominant process that
would preserve the spin polarization for small values of $k_z$ opening
up the possibility of using transport measurements to explore the
preferred spin channels according to the direction of propagation of
spin polarized currents along the quasi-1D
nanostructures.~\cite{Lelovsky}
In this way and by using the flexibility of manufacturing polytypic
nanowire (PNW) structures, we developed transport simulations in a NW
presenting regions with and without centers of bulk inversion asymmetry
that act as spin scatterers.
Thus, if carries are injected
along the $z$-axis of a  PNW
with certain incident energy, the peculiar symmetry of excited
states in the zincblende cylindrical NW region allows only the
transmission of a high degree of spin-polarized current according to
the propagation direction. A dominant spin-up or spin-down character of the current depends on
the sign of $k_z$ and the current density is
directly linked to the interplay between the wire radius and the
wavevector.

Likewise, motivated by the successful confirmation of surface effects on transparency modulation of 1D twin-plane based superlattices
(TWPSL), the Multicomponent Scattering Approach (MSA)~\cite{Tsu, SADiago2012} was applied to model and prescribe appealing tunneling
events for both uncoupled light-(lh) and heavy-holes (hh) through these NWs. Phase-time calculations show several anomalous
transport properties for each kind of charge carriers and reflect, as expected, the mini-band spectrum for 1D-TWPSL.~\cite{VLR-GEM11}
For instance, a clear hallmark of negative values of the phase time will be described. Undoubtedly, such evidences are known and were reported before for
electrons, holes, and twin-photons in typical $III-V$ layered structures~\cite{Vetter, Dolling, Wang, Chen, Muga} but, as far as we know,
they are novel events for TWPSL wires. Additionally, under certain circumstances, oscillating regions were detected for both carriers as
they trespass the $n$ barriers of the 1D-TWPSLs. This slightly recalls the Ramsauer-Townsend effect.~\cite{Townsend, Ramsauer, SADiago2012}

By heeding these two guidelines, the rest of the paper is organized as follows. The next two sections present the theoretical methods used to tackle the
transport simulations through spin-orbit interaction centers and twin-planes lattices. Afterwards, the main results are presented followed by the concluding remarks.

\section{Dresselhaus spin-orbit effect in polytypic nanowires}
\label{DSOI}

We simulate various transport properties in a NW presenting regions with and without centers of inversion
asymmetry (BIA) introduced by the Dresselhaus SO potential as sketched in Fig.~\ref{Fig1}~(a). The wavevector
has to be considered as an operator which fulfills the temporal inversion symmetry and the single-particle Hamiltonian for the conduction
band of a NW, including Dresselhaus SO interaction (SOI), can be written as $\mathcal{H}=\mathcal{H}_{0}+\mathcal{H}_{D}$, where
$\mathcal{H}_{0}=[\hbar^{2}\mathbf{\hat{k}}^{2}/2\mathit{m}^{\ast} + V( \mathbf{r,}z)]I$ is the Hamiltonian for uncoupled spin-up and
spin-down states, $I$ is the $2 \times 2$ identity matrix, and $V(\mathbf{r},z)$ is the total spatial confinement with lateral
$\mathbf{r}=(x,y)$ and longitudinal $z$ along the wire axis for the set of cartesian coordinates. The effective BIA Hamiltonian can be
separated into three contributions,
\begin{equation}
\mathcal{H}_{\mathcal{D}}=\mathcal{E}_{0}a^{3}\left(
\begin{array}{cc}
\mathcal{H}_{2\mathcal{D}} & \mathcal{H}_{1\mathcal{D}}+\mathcal{H}_{3 \mathcal{D}} \\
\mathcal{H}_{1\mathcal{D}}^{\dag }+\mathcal{H}_{3\mathcal{D}}^{\dag } & - \mathcal{H}_{2\mathcal{D}}
\end{array}
\right) ,
\label{eq2}
\end{equation}
where $\mathcal{H}_{1\mathcal{D}}=-\gamma
_{D}\hat{k}_{z}^{2}\hat{k}_{-}$,
$\mathcal{H}_{2\mathcal{D}}=-\frac{1}{2}\gamma _{D}\hat{k}_{z}\left(
\hat{k}_{-}^{2}+\hat{k}_{+}^{2}\right)$, and
$\mathcal{H}_{3\mathcal{D}}=-\frac{1}{8}\gamma_{D}\left\{\hat{k}_{+},(\hat{k}_{+}^{2}-\hat{k}_{-}^{2})\right\}$.
They are, respectively, the linear, the quadratic, and the cubic
Dresselhaus SOI contributions written in terms of operators
$\hat{k}_{\pm}$ multiplied by the dimensionless parameter
$\tilde\gamma=\gamma_D/(a^{3}\mathcal{E}_{0})$, and with
$\mathcal{E}_{0}=\hbar ^{2}/(2\mathit{m}^{\ast }a^{2})$ being the
energy scale for the NW confinement inside the cylindrical region.
The Schr\"{o}dinger equation for $\mathcal{H}$ is separable in the
entire level spectrum as a product of purely in-plane localized
function of the radial coordinate $\mathbf{r}$ times the free motion
function along the $z-$axis, and can be written as,
\begin{widetext}
\begin{eqnarray}
\left(
\begin{array}{cc}
 (\mu _s^{2}-\frac{\partial^2 }{\partial z^2})\delta _{ss'}+i  h_{ss'} \beta(z)\frac{\partial}{\partial z}  &   - j_{s,s^{\prime }}^{+}\beta(z)+ g_{ss'}^{-}\beta(z)\frac{\partial^2}{\partial z^2} \\
 - j_{s,s^{\prime }}^{-}\beta(z)+ g_{ss'}^{+}\beta(z)\frac{\partial^2}{\partial z^2} & (\mu _s^{2}-\frac{\partial^2 }{\partial z^2})\delta _{ss'}-i h_{ss'}\beta(z)\frac{\partial}{\partial z}\\
\end{array}
\right) \left(
\begin{array}{c}
 f^{ss'}_{\alpha\alpha}(z) \\
 f^{ss'}_{\alpha'\alpha}(z) \\
\end{array}
\right) = \left(
\begin{array}{cc}
 E & 0 \\
 0 & E \\
\end{array}
\right) \left(
\begin{array}{c}
 f^{ss'}_{\alpha\alpha}(z) \\
 f^{ss'}_{\alpha'\alpha}(z) \\
\end{array}
\right).
 \label{eq3}
\end{eqnarray}
\end{widetext}
Here, the index $s$ represents the set of quantum
numbers $(n,m)$, ordered by increasing values of energy ($E_{n,m}$)
with $\mu_{n,m}=\sqrt{(E_{0}/\mathcal{E}_{0}-a^2k_{z}^{2})}$ being
the $m^{th}$-zero of the $n^{th}$-order Bessel function,
$J_{n}\left( \mu_{n,m}\right)=0$, and with the index $\alpha$
accounting for the spin-up and spin-down states. The innermost
matrix elements are $g_{ss'}^{\pm }=\left\langle s\left\vert
\hat{k}_{\pm }\right\vert s^{\prime }\right\rangle $,
$h_{ss'}=\frac{1}{2}\left\langle s\left\vert
\hat{k}_{+}^{2}+\hat{k}_{-}^{2}\right\vert s^{\prime}\right\rangle$,
and $j_{s,s^{\prime }}^{\pm}=\frac{1}{8}\left\langle s\left\vert
\{\hat{k}_{\pm }^{2},\left( \hat{k}_{+}^{2}-\hat{k}_{-}^{2}\right)
\}\right\vert s^{\prime }\right\rangle $. The Dresselhaus
coefficient, as function of position along the wire axis, can be
emulated as
\begin{equation}
\label{} \beta(z)=\frac{\tilde\gamma}{2 \text{erf}(\ell )}
[\text{erf}(z+\ell )-\text{erf}(z-\ell )],
\end{equation}
where $\text{erf}(z)$ is the error function and $2\ell$ is the width of the resulting quasi square region with finite first derivative, illustrated in
Fig.~\ref{Fig1}~(b). Note that the method presented here can afford any kind of potential regardless its spatial shape. After
symmetrization of the Dresselhaus Hamiltonian, in the dimensionless form, one reaches a nonlinear set of equations, written as
\begin{eqnarray}\label{nonl1}
&-f^{''ss'}_{\alpha\alpha}(z)+ i \beta(z) \tilde h f^{'ss'}_{\alpha\alpha}(z)+ \beta(z) \tilde g^- f^{''ss'}_{\alpha'\alpha}(z) \nonumber \\
&+ \beta '(z) \tilde g^- f^{'ss'}_{\alpha'\alpha}(z)
= (\mathit{k}^2+\mu_s^2) f^{ss'}_{\alpha\alpha}(z)  \\
&- \frac{1}{2}\left(i \beta'(z)\tilde h f^{ss'}_{\alpha\alpha}(z)+\left(\beta''(z) \tilde g^- +2 \tilde j^+\beta(z)\right)
f^{ss'}_{\alpha'\alpha}(z)\right) \nonumber
\end{eqnarray}
and
\begin{eqnarray}\label{nonl2}
&-f^{''ss'}_{\alpha'\alpha}(z)- i \beta (z) \tilde h f^{'ss'}_{\alpha'\alpha}(z)+ \beta(z) \tilde g^+ f^{''ss'}_{\alpha\alpha}(z) \nonumber \\
&+ \beta '(z) \tilde g^+ f^{'ss'}_{\alpha\alpha}(z)
=(\mathit{k}^2+\mu_s^2 ) f^{ss'}_{\alpha'\alpha}(z)  \\
&+ \frac{1}{2}\left(i \beta '(z) \tilde h f^{ss'}_{\alpha'\alpha}(z)-\left(\beta''(z) \tilde g^+ -2 \tilde j^-\beta(z)\right)
f^{ss'}_{\alpha\alpha}(z)\right),\nonumber
\end{eqnarray}
where $k=a k_z$, $\tilde g^{\pm}=a g_{ss'}^{\pm}$, $\tilde
h=a^2h_{ss'}$ and $\tilde j^{\pm}=a^3 j_{ss'}^{\pm}$. In order to
solve this set of equations we have used the Reverse Runge-Kutta
method as shortly described below.

\subsection{Reverse Runge-Kutta}

For two spin channels and a single subband $s=s'$, one can write the wavefunction in each of the three regions delimited in Fig.~\ref{Fig1}~(b) as,
\begin{eqnarray}
\label{}
\Psi_{\text{I}}=\left(
\begin{array}{cc}
 e^{-i z k} R_{\alpha\alpha}+e^{i z k} \\
 e^{-i z k} R_{\alpha'\alpha} \nonumber
\end{array}
\right),
\end{eqnarray}

\begin{eqnarray}
\label{}
\Psi _{\text{II}}=\left(
\begin{array}{cc}
 \psi _{\alpha\alpha} \\
 \psi _{\alpha'\alpha}
\end{array}
\right),\;\mbox{and} \;\; \Psi _{\text{III}}= \left(
\begin{array}{c}
 e^{i z k} T_{\alpha\alpha} \\
 e^{i z k} T_{\alpha'\alpha} \nonumber
\end{array}
\right),
\end{eqnarray}
respectively. From the right interface boundary condition,
$\Psi_{\text{II}}(L)=\Psi_{\text{III}}(L)$, one can write
\begin{equation}
\label{}
\text{F} _{\text{III}}(L)=\left(
\begin{array}{c}
 \frac{\psi _{\alpha\alpha}(L)}{T_{\alpha\alpha}} \\
 \frac{\psi _{\alpha'\alpha}(L)}{T_{\alpha'\alpha}}
\end{array}
\right)  \equiv \left(
\begin{array}{c}
 e^{i z k} \\
 e^{i z k} \nonumber
\end{array}
\right), \; \mbox{and}
\end{equation}

\begin{equation}
\label{}
\text{F}_{\text{II}}(z)= \left(
\begin{array}{c}
 f^{ss}_{\alpha\alpha}(z) \\
 f^{ss}_{\alpha'\alpha}(z)
\end{array}
\right)\equiv\left(
\begin{array}{c}
 \frac{\psi _{\alpha\alpha}(z)}{T_{\alpha\alpha}} \\
 \frac{\psi _{\alpha'\alpha}(z)}{T_{\alpha'\alpha}}
\end{array}
\right), \nonumber
\end{equation}
where $\text{F}_i$ is the dispersion region function. Hence,
$\text{F}_{\text{II}}(L)=\text{F}_{\text{III}}(L)$, and
$\text{F}^\prime_{\text{II}}(L)=\text{F}^\prime_{\text{III}}(L)$
are, respectively,
\begin{equation}
\label{boundayright}
\left(
\begin{array}{c}
 f^{ss}_{\alpha\alpha}(L) \\
 f^{ss}_{\alpha\alpha'}(L)
\end{array}
\right)=\left(
\begin{array}{c}
 e^{i L k} \\
 e^{i L k}
\end{array}
\right),\; \mbox{and}\left(
\begin{array}{c}
 f^{'ss}_{\alpha\alpha}(L) \\
 f^{'ss}_{\alpha'\alpha}(L)
\end{array}
\right)=\left(
\begin{array}{c}
 i e^{i L k} k \\
 i e^{i L k} k
\end{array}
\right).
\end{equation}

Following the reversal procedure for the left boundary
condition, where $\Psi_{\text{I}}(-L)=\Psi_{\text{II}}(-L)$), one
can address the unknown function $\text{F}_{\text{I}}$ as,
\begin{equation}
\label{}
\text{F}_{\text{I}}(-L)=\left(
\begin{array}{c}
 \frac{\psi_{\alpha\alpha}(-L)}{T_{\alpha\alpha}} \\
 \frac{\psi_{\alpha'\alpha}(-L)}{T_{\alpha'\alpha}}
\end{array}
\right), \; \mbox{and} \nonumber
\end{equation}

\begin{equation}
\label{}
\text{F}_{\text{I}}(z)= \left(
\begin{array}{c}
 \frac{e^{i z k} R_{\alpha\alpha}+e^{-i z k}}{T_{\alpha\alpha}} \\
 \frac{e^{i z k} R_{\alpha'\alpha}}{T_{\alpha'\alpha}} \\
\end{array}
\right). \nonumber
\end{equation}
Therefore, from the left boundary condition,
$\text{F}_{\text{I}}(-L)=\text{F}_{\text{II}}(-L)$, and
$\text{F}^\prime_{\text{I}}(-L)=\text{F}^\prime_{\text{II}}(-L)$,
along with the current conservation,
 $\left|
T_{\alpha\alpha}(\mathit{k})\right|^2+\left|
R_{\alpha\alpha}(\mathit{k})\right|^2+ \left|
T_{\alpha'\alpha}(\mathit{k})\right|^2+\left|
R_{\alpha'\alpha}(\mathit{k})\right|^2=1$, one is able to define the
transmission ($T$) and reflection ($R$) coefficients in terms of the
spin components. Thus,
\begin{equation}
\label{}
T_{\alpha\alpha}(k)=\frac{2 \mathit{k}}{\mathit{k}
f^{ss}_{\alpha\alpha}(-L)-i f^{'ss}_{\alpha\alpha}(-L)},
\end{equation}
\begin{equation}\label{}
R_{\alpha\alpha}(k)=\frac{\mathit{k} f^{ss}_{\alpha\alpha}(-L)+i
f^{'ss}_{\alpha\alpha}(-L)}{\mathit{k} f^{ss}_{\alpha\alpha}(-L)-i
f^{'ss}_{\alpha\alpha}(-L)},
\end{equation}
\begin{equation}\label{}
T_{\alpha'\alpha}(k)=k \sqrt{\frac{1-\left|
R_{\alpha\alpha}(\mathit{k})\right|^2-\left|
T_{\alpha\alpha}(\mathit{k})\right|^2}{k^2+\left|f^{ss}_{\alpha'\alpha}(-L)\right|^2}},
\; \mbox{and }
\end{equation}
\begin{equation}\label{}
R_{\alpha'\alpha}(k)= \sqrt{\frac{1-\left|
R_{\alpha\alpha}(\mathit{k})\right|^2-\left|
T_{\alpha\alpha}(\mathit{k})\right|^2}{k^2+\left|f^{ss}_{\alpha'\alpha}(-L)\right|^2}}\left|f^{'ss}_{\alpha'\alpha}(-L)\right|
\end{equation}
where the $f(-L)$ is obtained employing the right boundary
conditions, Eq.~(\ref{boundayright}), along with the $z$-component
Schr\"{o}dinger Eqs.~(\ref{nonl1}) and (\ref{nonl2}).  Similar
expressions are obtained extending this method for multiple
subbands.

\section{Phase-time in twin-plane superlattices}
\label{TMM}

In turn, for the simulation of the tunneling properties in the twin plane NW we have used the transfer matrix framework assuming a 1D-TWSL, represented in Fig.~\ref{Fig1}(c) with alternated layers whose quantum heterogeneity, due to strain field modulation, is revealed along the $z$-axis in Fig.~\ref{Fig1}(d). The envelope function (EF) coefficients at both TWSL extremal slabs, are bounded as
\begin{eqnarray}
     \label{for:TMcel}
            \begin{bmatrix}
       \begin{array}{c}
        A_{1} \\
        B_{1}
       \end{array}
      \end{bmatrix}
        =
        \mathbf{M}_{1}(z_{0})^{-1}\cdot\mathbf{M}_{2}(z_{2})\ldots\cdot\mathbf{M}_{4}(z_{4})^{-1}\cdot
        \\ \nonumber
        \mathbf{M}_{5}(z_{4})
      \begin{bmatrix}
       \begin{array}{c}
         A_{5} \\
         B_{5}
       \end{array}
      \end{bmatrix}, \,\,\;\;\;\;\;\;
\end{eqnarray}
being
\begin{eqnarray*}
        \mathbf{M}_{i} =
         \begin{pmatrix}
          \begin{array}{cc}
           e^{\imath k_{i}z_{i}} & e^{-\imath k_{i}z_{i}} \\
           \imath v_{i}e^{\imath k_{i}z_{i}} & i v_{i} e^{-\imath k_{i}z_{i}}
          \end{array}
         \end{pmatrix},
\end{eqnarray*}
the transfer matrix. The EF is given by
\begin{equation}
        \Psi_{i}(z) = A_{i} e^{\imath k_{i}z_{i}} + B_{i} e^{-\imath k_{i}z_{i}},
\end{equation}
and its weighted derivative, as
\begin{equation}
        \frac{1}{m_{i}}\Psi_{i} p(z) = \frac{\imath k_{i}}{m_{i}} \left( A_{i} - B_{i} \right),
\end{equation}
at both layers $I$ and $V$ of the 1D-TWPSL. [see Fig.~\ref{Fig1}(\textbf{d})]. The boundary conditions for scattering quantities can be
set by assuming an impinging mixing-free particle (quasi-particle) traveling from the left layer $I$, represented in Fig.~\ref{Fig1}(d),
 \begin{equation}
  \label{for:IC}
  \left\lbrace
    \begin{array}{lcl}
      A_{1} = 1 &;& \; B_{1} = T \\
      A_{5} = R &;& \; B_{5} = 0
   \end{array},
  \right.
\end{equation}
where $T$ and $R$ stand for transmission and reflection amplitudes,
respectively. Next, we establish a correlation between quantities
(\ref{for:TMcel}) and (\ref{for:IC}), as
\begin{eqnarray}
  \label{for:TMcel2}
  \begin{bmatrix}
     \begin{array}{c}
      1 \\
      R
     \end{array}
  \end{bmatrix}
  = \mathbf{M}(z_{l}, z_{r})\cdot
    \begin{bmatrix}
     \begin{array}{c}
      T \\
      0
     \end{array}
    \end{bmatrix}
  =  \\ \nonumber \begin{pmatrix}
       \begin{array}{cc}
        M_{11}(z_{l}, z_{r}) & M_{12}(z_{l}, z_{r}) \\
        M_{21}(z_{l}, z_{r}) & M_{22}(z_{l}, z_{r})
       \end{array}
     \end{pmatrix}\cdot
     \begin{bmatrix}
      \begin{array}{c}
       T \\
       0
      \end{array}
    \end{bmatrix},
\end{eqnarray}
which straightforwardly lead us to
\begin{equation}
 \label{for:SC}
     \begin{array}{lcl}
        T_{ij} = T_{\Re_{ij}} + \; \imath T_{\Im_{ij}} = \left[ M_{11}(z_{l},
            z_{r})\right]^{-1} \\
        R_{ij} = R_{\Re_{ij}} + \; \imath R_{\Im_{ij}} = \frac{M_{21}(z_{l},
            z_{r})}{M_{11}(z_{l}, z_{r})},
    \end{array}
\end{equation}
the amplitude matrices for transmission and reflection,
respectively. Next, we obtain from $|T_{ij}|^{2}$ and $|R_{ij}|^{2}$
the scattering probabilities for the transmitted and reflected
fluxes, respectively. For the calculation of the tunneling phase-time, a
simple procedure is followed. Firstly, we calculated the
transmission phase
\begin{eqnarray}
 \label{for:Fase}
   &\theta_{ij} = \arctan\left\{\frac{T_{\Im_{ij}}}{T_{\Re_{ij}}}\right\}+\frac{\pi}{2}\left(1-\frac{T_{\Re_{ij}}}{|T_{\Re_{ij}}|}\right)=
   \\
   &\arctan\left\{\frac{\Im\left(M_{11}(z_{l}, z_{r})\right)}{\Re\left(M_{11}(z_{l}, z_{r})\right)}\right\}+\frac{\pi}{2}\left(1-\frac{\Re\left(M_{11}(z_{l}, z_{r})\right)}{|\Re\left(M_{11}(z_{l}, z_{r})\right)|}\right),
    \nonumber
\end{eqnarray}
and afterwards, we derive the $n$-cell tunneling phase-time\cite{SADiago2012,LCRP06} $\left(\tau_{ij}\right)_{n} =
\hbar\frac{\partial}{\partial \mathcal{E}}\theta_{ij}$, assuming left-incoming particle (quasi-particle)
\begin{equation}
 \label{for:FaseT}
 \left(\tau_{ij}\right)_{n}
   =  \frac{\hbar\,\Re^{2}(T_{n_{ij}})}{\Re^{2}(T_{n_{ij}})+
        \Im^{2}(T_{n_{ij}})}
      \left\{\frac{\partial}{\partial \mathcal{E}}
        \left[\frac{\Im(T_{n_{ij}})}{\Re(T_{n_{ij}})}\right]
      \right\}\,.
\end{equation}

\section{Results and discussion}

The formalism developed in Sec.~\ref{DSOI} allows the study of spin
transport of PNWs, where changes in the zincblende region width, $w$,
the NW diameter, $\phi$, the incident energy, $\mathcal{E}$, and
the SOI strength, $\beta$, can be used as tuning parameters. The
presented results correspond to an incident spin polarized current
injected on the wurtzite region of a InSb PNW, where
$\beta=760.1$~eV$\cdot$\AA$^{3}$. We will see that the SOI
within the zincblende region gives rise to a small net polarization.
Projecting it to several sequential regions, it is possible to realize
spin filters and modulate the densities of spin-polarized current flowing in both
parallel and antiparallel directions of the NW.

\begin{figure}[h]
\includegraphics[scale=1.0]{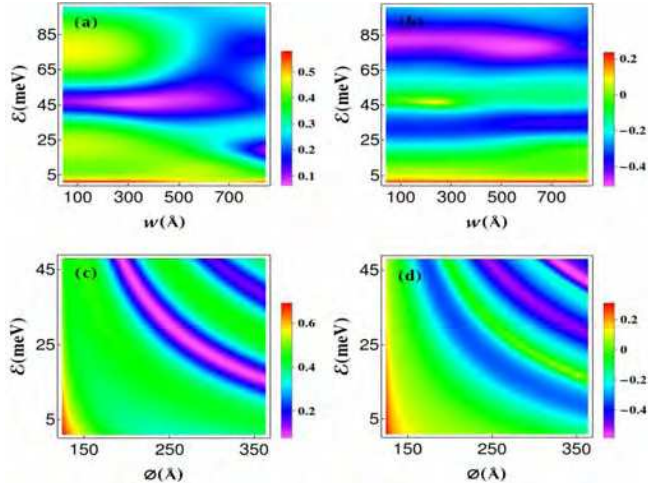}
\caption{Persistence contour maps of polytypic InSb quantum wire for
the first band where $\beta=760.1$~eV$\cdot$\AA$^{3}$. Panels (a)
and (b) show the transmission and reflection as function of the
barrier width and energy for a wire with radius 101~\AA. In
panels (c) and (d) we show transmission and reflection as function
of the wire diameter and energy.}
\label{fig5}
\end{figure}

In order to characterize the system response to the injection of a
spin-unpolarized superposition of spin-up and spin-down currents, we
calculate the spin persistence ratio as defined by
$[|T_{\alpha\alpha}|^{2} +
|T_{\alpha'\alpha'}|^{2}-(|T_{\alpha'\alpha}|^{2}+
|T_{\alpha\alpha'}|^{2})]/T$, where $T = |T_{\alpha\alpha}|^{2} +
|T_{\alpha'\alpha'}|^{2}+ |T_{\alpha'\alpha}|^{2}+
|T_{\alpha\alpha'}|^{2}$. Figure~\ref{fig5} shows the color code maps
of characteristic persistence vs. incident energy, $\mathcal{E}$, and $w$, and
vs. $\mathcal{E}$ and $\phi$. Note, that the persistence is always
positive, while characterizing the transmitted current, but does not
reach the unity (Figs.~\ref{fig5}~(a) and (c)). This positiveness
refers to a signature of almost perfect spin-preserving channel, as
expected due to the non-degenerated ground state,~\cite{Lelovsky}
but there are oscillating regions which evolve with changes in
$\mathcal{E}$, $w$, and $\phi$. On the other hand, from the
characterized reflected current it is possible to observe that the
persistence goes from spin-preserving to spin-reversing (negative
values) channels. The spin-reversing regions are identified by
negative numbers and also evolve with changes in $\mathcal{E}$, $w$,
and $\phi$. Despite the difference of spin-preserving and
spin-reversing channels, the oscillation pattern of the reflection
coefficient is doubled in comparison to the transmitted one and this
should be expected once the current crosses twice the region of
Dresselhaus SOI.~\cite{coupler} Moreover, the evolution of the
oscillations goes with both $\mathcal{E}$ and $\phi$, so that
$\mathcal{E} \phi=$ constant, as shown in panels (c) and (d) of
Fig.~\ref{fig5}, due to the modulation of the energy levels with
$\phi$, Eq.~(\ref{eq3}).

\begin{figure}[h]
\includegraphics[scale=1.0]{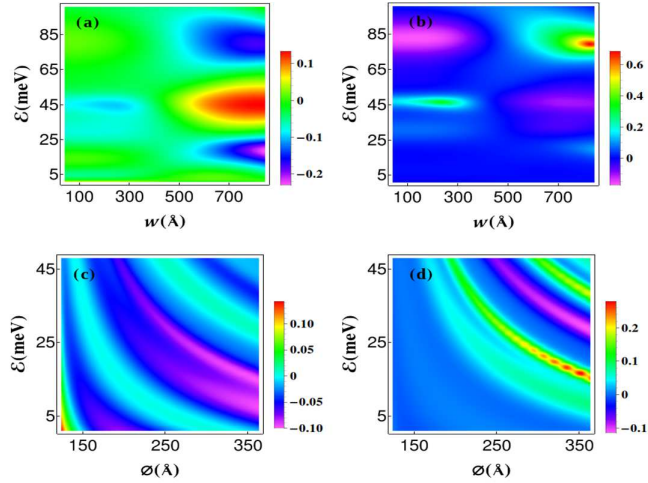}
\caption{Polarization contour maps of polytypic InSb NW for the
first subband $\beta=760.1$~eV$\cdot$\AA$^{3}$. Transmission (a) and
reflection (b) as function of the barrier width and incident energy
for a wire with radius of 101~\AA. Transmission (c) and reflection
(d) as function of the wire diameter and incident energy.}
\label{fig4}
\end{figure}

When considering excited states in the transport, one can expect a
small net spin polarization for the current. This is due to the fact
that for transport along the $z$-direction, the Dresselhaus SOI will
produces the precession of both spin-components but in different
ways.~\cite{Lelovsky} The degree of spin-polarization is defined by
$[|T_{\alpha\alpha}|^{2} +
|T_{\alpha'\alpha}|^{2}-(|T_{\alpha'\alpha'}|^{2} +
|T_{\alpha\alpha'}|^{2})]/T$. Figure~\ref{fig4} shows the
polarization as function of $\mathcal{E}$ and $w$ in panels (a) and
(b), and as a function of $\mathcal{E}$ and $\phi$, in panels (c) and
(d).

Observe that the transmission displayed in Fig.~\ref{fig4}~(a) shows well defined polarization regions which evolve with both $\mathcal{E}$
and $w$. Around $\mathcal{E}=5$~meV and $w=150$~\AA, a smaller percentage of spin-down polarization is evident, as well as for
$\mathcal{E}=45$~meV. Looking at $w=800$~\AA, one can see an opposite spin-polarization for $\mathcal{E}=45$~meV. Moreover, it is clear that
at this width $w$, a modulation of the polarization takes place, from spin-down to spin-up character, by varying $\mathcal{E}$. This last
hallmark is also revealed in the reflection panel, shwon in Fig.~\ref{fig4}~(b), but with opposite polarization features. Additionally, Figs.~\ref{fig4}~(c)
and (d) show the same behavior of Figs.~\ref{fig5}~(c) and (d), with oscillations occurring by varying both $\mathcal{E}$ and $\phi$, such that
$\mathcal{E} \phi=$constant. These results indicate the possibility of modulation of the polarization by selecting the incident energy of
carriers, the NW wire width, the zincblend region width, and the SO potential.

\begin{figure}[h]
\includegraphics[scale=1.0]{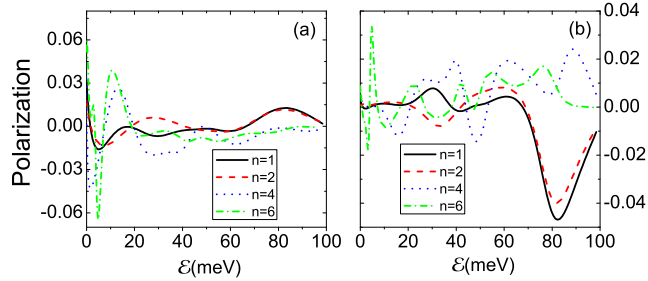}
\caption{Polarization of polytypic InSb NW with multiple wurtzite
regions, n, where $\beta=760.1$~eV$\cdot$\AA$^{3}$. (a) Transmission
and (b) reflection as function of the energy for a wire with radius
of 101~\AA.} \label{fig6}
\end{figure}

The presented results for the spin persistence and polarization refer only to a configuration of one zincblend region between two wurtzite
layers, as seen in Fig.~\ref{Fig1}~(b), and this was enough to characterize a small percentage of spin flux modulation.
Increasing the number of layers containing the zincblende regions, one can expect an increase of the polarization. Figure~\ref{fig6} shows
the transmission, (a), and reflection, (b), polarizations for multiple wurtzite/zincblende/wurzite regions. As can be noted, an increasing
number of layers leads to increasing spin filtering and enhancement of the final polarization.

Along with the polytypism, the growth conditions may lead to controlled stacking faults appearance and the synthesis of twin plane superlattices. For these later, as represented in as Fig.~\ref{Fig1}(c), we have proposed a transport analysis of the tunneling phase-time for both lh and hh. Here, the results for
electrons were not shown due to similarities with the lh transport results.~\cite{VLR-GEM11}

\begin{figure}[h]
\includegraphics[scale=1]{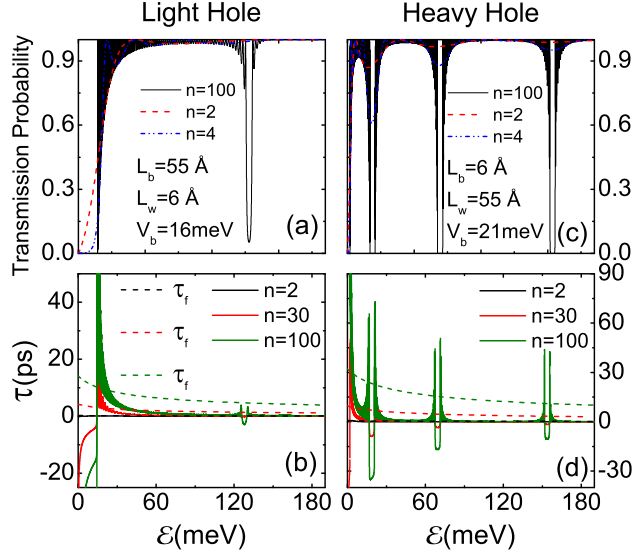}
\caption{Calculated transmission probability and phase-time ($\tau_{n}$) for various sequences of twin-plane cells, $n$, in a 1D-TWPSL as
a function of incident energy ($\mathcal{E}$): (a) and (b) Transmission probability, $\tau$ and quasi-classical free motion time
($\tau_{f}$), respectively, for light-holes (lh). (c) and (d) Transmission probability, $\tau$ and $\tau_{f}$, respectively, for
heavy-holes (hh).} \label{fig2}
\end{figure}

Using the formalism of Sec.~\ref{TMM}, the mini-band spectrum for holes tunneling through ($n = 2-100$) 1D-TWPSL is depicted in
Fig.~\ref{fig2} as $\tau_{n}$ (\ref{for:FaseT}) and evolves with the carrier incident energy, $\mathcal{E}$. As expected, this simulation
agrees with an previous direct calculation of the 1D-TWPSL spectrum~\cite{VLR-GEM11} and responds to the mini-band spectrum of the 1D-TWPSL,
\textit{i.e.} it reproduces accurately the quasi-stationary holes states of the embedded 1D-TWPSL quantum wells. We found the double
resonant tunneling barrier (DBRT) curve ($n=2$, red dashed line) to represent a lower bound for the 1D-TWPSL mini-band spectrum as had
been reported for electrons~\cite{PPP00} and for holes~\cite{SADiago2012} in two-dimensional systems (see Fig.~\ref{fig2} (a) and
(c)). Appealing regularity for energy sections has been found, where $\tau_{100} <\tau_{30} < \tau_{2} <0$ (see dips in Figs.~\ref{fig2}(b)
and (d)), where the transmission vanishes and, thereby, coincides with the mini-band forbidden regions of the system (Figs.~\ref{fig2}(a)
and (c)). In other words, this abnormal behavior for holes is directly connected to a large backscattering
($|T(\mathcal{E})|^{2}\simeq0$).~\cite{Hauge, Vetter, Dolling} The later could be due to the fact that negative values of $\tau$ are not
only related to the incident wave from left but also might be a combination of both incident and accumulatively reflected waves from the
1D-TWPSL quantum barriers traveling towards the left-hand side.~\cite{Dolling} One may notice that the quasi-classical free motion time,
$\tau_{f} = \frac{nL_{z}m_{eff}}{\hbar k}$, which characterizes the temporal scale for a particle with mass $m_{eff}$ to travel through a
none-scattering space region of dimension $n$ times the thickness $L_{z}$,~\cite{SADiago2012} can be straightforwardly compared with
Eq.~(\ref{for:FaseT}). Indeed, in our case, a striking quality arises whenever $\left(\tau_{f}\right)_{2,30,100}
> \tau_{2,30,100}$, for the mini-gap regions,
respectively (Fig.~\ref{fig2} (b) and (d)). Therefore, for incoming hh states, with energy in the neighborhood of $20$~meV,
$\left(\tau_{f}\right)_{100} - \tau_{100} \approx 60$~ps, while for incoming lh with incident energy $\sim 120$~meV,
$\left(\tau_{f}\right)_{100} - \tau_{100} \approx 15$~ps. This earlier arrival time for both hh and lh, previously predicted for
electrons~\cite{PPP00} and holes~\cite{SADiago2012} that tunnel through a two-dimensional DBRT and a SL, suggests a faster passage of holes trough a
1D-TWPSL, under specific conditions. The regularity of more speedily passage of hh, compared to that of lh,~\cite{SADiago2012} is clearly
preserved in our numerical simulation. Importantly, only at low energies ($\sim 5$ meV for hh, and $\sim[14-23]$ meV for lh) $\tau_{30}$
approaches $\left(\tau_{f}\right)_{30}$ from upper side down, as detected for electrons~\cite{PPP00} and holes in two-dimensional
SL.~\cite{SADiago2012} Additionally, it seems to be satisfied that $\left(\tau_{f}\right)_{2,30,100} > \tau_{2,30,100}$ for the rest of
impinging energy values in the transparent region, contrary to the two-dimensional case.~\cite{SADiago2012}

\begin{figure}[h]
\includegraphics[scale=1.0]{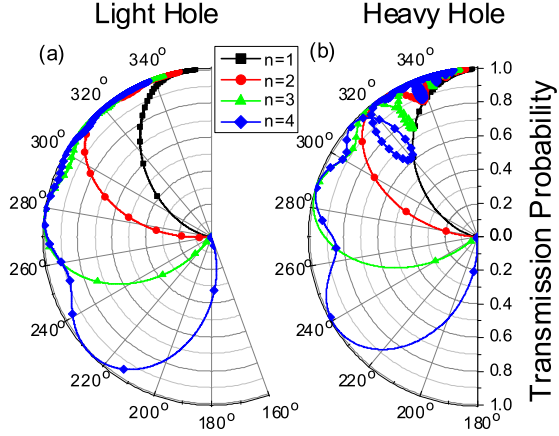}
\caption{Representation of the transmission amplitude through different numbers of cells, $n$: $T_{ij} = T_{\Re_{ij}} + \imath
T_{\Im_{ij}}$, in the complex plane for increasing incident energy. (a) lh and (b) hh. Note that both diagrams have been rotated for
clarity.} \label{fig7}
\end{figure}

In order to spread some light and try to elucidate some features of the negative phase-time (eq.~(\ref{for:FaseT})) values, one can see in
Fig.~\ref{fig7} the transmission amplitude, $T_{ij} = T_{\Re_{ij}} + \imath T_{\Im_{ij}}$, in the complex plane for different numbers of
cells, $n$. Here, $\theta_{ij}(\mathcal{E})$ has been taken as the angle and $|T(\mathcal{E})|^{2}$ as the radius. As can be seen, the major
signature of the curves is their clockwise direction evolution, which can be directly related with positive phase-time values. However,
counterclockwise segments clearly appear for hh with $n=3$ (green-triangles) and $n=4$ (blue-diamonds), characterized by loops
in Fig~\ref{fig7}(b). Interestingly, these loops seems to be modulated by the $n=1$ (black-squares) curve as a lower bound,
Fig~\ref{fig7}(b). The last could be a certain resemblance of a similar average behavior, detected for electrons~\cite{PPP00} and
holes~\cite{SADiago2012} in two-dimensional systems. The counterclockwise direction of $\theta_{ij}(\mathcal{E})$ is directly related with negative
values of the phase-time. For lh, in Fig.~\ref{fig7}(a), this occurs when $|T(\mathcal{E})|^{2}\simeq0$ for $n \geq 4$ and for a large number
of cells. One can find this negative values of $ \tau $ for hh when $|T(\mathcal{E})|^{2}\simeq0$ for $n\geq3$, in Fig.~\ref{fig7}(b). Despite
some criticism in the specialized literature, clear hallmarks of negative values for $\tau$ are reported, where any numerical artifact has
been accurately excluded. In spite of the interpretation proposed above, for this counterintuitive topic, a conclusive robust theoretical
modeling remains a puzzle. Such evidences are novel events for TWPSL NWs, although they are well established on other
systems.~\cite{Vetter, Dolling, Wang, Chen, Muga}

\begin{figure}[h]
\includegraphics[scale=1.0]{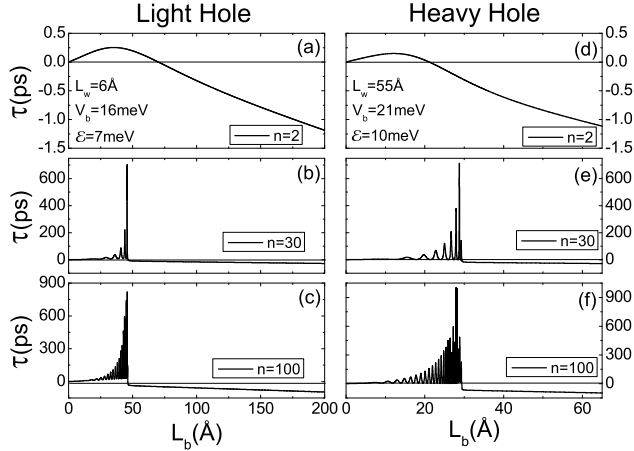}
\caption{Calculated $\tau_{n}$ for various sequences of twin-plane cells in a 1D-TWPSL as a function of the barrier width ($L_{b}$). For
lh: (a), (b) , and (c). For hh: (d), (e), and (f).} \label{fig3}
\end{figure}

Figures~\ref{fig3}(b) and (c) display the calculated $\tau_{n}$, eq.~(\ref{for:FaseT}), for different sequences of TWP cells caused by the
strain effect, as a function of the barrier width, $L_{b}$. Oscillating regions are detected for both hh and lh, as they trespass the $n
= 30, 100$ 1D-TWPSLs with low values of energy, $\mathcal{E} < 0.5V_{b}$ [see Fig.~\ref{fig3}~(b),~(c),~(e),and~(f)]. Accumulative barrier
interferences with evanescent hole states have been assigned as the possible origin of these oscillations. This slightly recalls the
Ramsauer-Townsend effect.~\cite{SADiago2012} This behavior fades away abruptly at $L_{b} \approx 50$ {\AA} for lh whereas for hh is
observed at $L_{b} \approx 30$ {\AA}. It can be argued that beyond a critical value of barrier width, the wavefunction penetration length
becomes negligibly small and, thereby, the $|T(\mathcal{E})|^{2}$ tends to vanish while $\tau_{n} < 0$ values arise. None of such
oscillations has been found for the DBRT case ($n = 2$).

\section{Conclusion}

In summary, motivated by well established methods for manufacturing polytypic and twin-plane NWs, we have proposed two different carrier transport
studies for the characterization of these systems. The first one was the spin transport simulation on a polytypic NW structure presenting
regions with (zincblende) and without (wurtzite) Dresselhaus SO potential. For this transport calculation we developed a new approach
based on the Reverse Runge-Kutta method. We have found that carries injected with certain energy along the $z$-axis of a TPNW, would allow the transmission of spin-up current along the $+z$-direction and spin-down current along the $-z$-direction, as due to
the peculiar symmetry of the excited states in the zincblende cylindrical region. Furthermore, the increasing number of layers in the
structure leads to increasing spin filtering ability as well as an enhancement of the final current polarization. The spin-polarized
current density can be directly linked to the interplay between the wire radius and width of the Dresselhaus SOI layer.

In turn, the second simulation uses the MSA and TMM methods to study the charge carriers passage through a strained 1D-TWPSL. As expected,
the calculation averaged out the mini-band spectrum for 1D-TWPSL. Amplitude oscillating regions were obtained for both hh and lh carrier
currents with low values of incident energy, thus evoking the Ramsauer-Townsend effect. Moreover, the phase-time $\tau_{n}$ assumes
negative values whenever the 1D-TWPSL properties reach a minimum or negligible transmission probabilities. This abnormal behavior is
directly connected to a large carrier backscattering at barrier regions of the finite superlattice.

In order to elucidate this effect, we have characterized the transmission amplitude in the complex plane where the major clockwise
signatures were directly related to positive values of  $\tau_{n}$, whereas the minor counterclockwise ones were connected to negative
values of the phase-time. This effect is only observed for structures with $n > 2$. We hope that these findings would stimulate
further search for the realization of spin filtering structures.

\begin{acknowledgements}
 The authors are grateful for financial support by the Brazilian Agencies CNPq, CAPES, FAPESP (Grant $\sharp$ 2012/02655-1), and the Academic Visiting Program of CLAF-PLAF of the Brazilian Physical Society. L.D-C is grateful for the hospitality of UFSCar Physics Department.
\end{acknowledgements}

\end{document}